\documentclass{aastex}
\usepackage{emulateapj5}

\newcommand{\kms}{km\thinspace s$^{-1}$}
\newcommand{\ergss}{ergs\thinspace s$^{-1}$}

\slugcomment{to be published in ApJ Letters}

\shorttitle{[OIII] 5007 Emission from the BH XRB in a NGC~4472 GC}
\shortauthors{Zepf et al.}

\begin{document}


\title{[OIII] 5007 Emission from the Black-Hole X-ray Binary in a NGC~4472
Globular Cluster\altaffilmark{1,2}}


\author{Stephen E. Zepf\altaffilmark{3}, Thomas J. Maccarone\altaffilmark{4}, 
Gilles Bergond\altaffilmark{3,5}, Arunav Kundu \altaffilmark{3}, 
Katherine L. Rhode\altaffilmark{6,7}, \& John J. Salzer\altaffilmark{6,7}}

\altaffiltext{1} {Based on observations performed at the European Southern 
Observatory Very Large Telescope on Cerro Paranal, Chile under
programs 72.B-0384 and 75.B-0516}
\altaffiltext{2} {Based on observations obtained with XMM-Newton, 
an ESA science mission with instruments and contributions directly 
funded by ESA Member States and NASA}
\altaffiltext{3}{Department of Physics \& Astronomy, Michigan State University,
East Lansing, MI 48824; e-mail: zepf,akundu@pa.msu.edu}
\altaffiltext{4}{School of Physics \& Astronomy, University of Southampton, 
Southampton, Hampshire SO17 1BJ; e-mail: tjm@astro.soton.ac.uk}
\altaffiltext{5}{Instituto de Astrof\'\i sica de Andaluc\'\i a, C/ Camino 
Bajo de Hu\'etor, 50, 18008 Granada, Spain; e-mail'' gbergond@caha.es}
\altaffiltext{6}{Department of Astronomy, Indiana University,
Bloomington, IN 47405; e-mail: rhode,slaz@astro.indiana.edu}
\altaffiltext{7}{Department of Astronomy, Wesleyan University,
Middletown, CT 06459}
%

\begin {abstract}
We present the discovery of [OIII] 5007 emission 
associated with the black hole X-ray binary recently identified 
in a globular cluster in the Virgo elliptical galaxy NGC~4472. 
This object is the first confirmed 
black-hole X-ray binary in a globular cluster.
The identification of [OIII] 5007 emission from the
black-hole hosting globular cluster 
is based on two independent fiber spectra obtained at the VLT with FLAMES, 
which cover a wavelength range of 5000-5800 \AA\ 
at a spectral resolution of about 6000.
In each of these spectra we find an emission line at 5031.2 \AA\
with an uncertainty of several tenths of an angstrom.
These are consistent with [OIII] 5007 emission at the $1475 \pm 7$ \kms\ 
radial velocity of the globular cluster previously determined
from an analysis of its absorption lines. This agreement
within the small uncertainties argues strongly in favor
of the interpretation of the line as [OIII] 5007 emission from
the black-hole hosting globular cluster. We also find
that the emission line most likely has a velocity width of
several hundred \kms. Such a velocity width rules out a 
planetary nebula explanation for the [OIII] 5007 emission 
and implicates the black hole as the source of the power 
driving the nebular emission.

\end{abstract}

\keywords{
galaxies:individual (NGC~4472) --- galaxies: star clusters --- 
globular clusters: general --- X-rays: binaries --- X-rays: galaxies: clusters}

\section{Introduction}

Maccarone et al.\ (2007, hereafter MKZR07) presented the discovery 
of a black-hole X-ray binary in a globular cluster in the Virgo
elliptical NGC~4472. 
This is the first unambiguous 
black-hole X-ray binary in any globular cluster.
This result 
is based on a 100 ks XMM-Newton observation of the Virgo elliptical
NGC~4472 and its globular clusters.
Specifically, MKZR07 identified an X-ray source in a 
spectroscopically confirmed globular cluster
of the Virgo elliptical galaxy NGC~4472 with an X-ray luminosity
of $4 \times 10^{39}$ \ergss\ for which the observed count rate
varied by a factor of seven over a few hours. 
The high X-ray luminosity, about ten times $L_{\rm Edd}$ for a
neutron star, rules out any single object other than
a black hole in the old stellar population of the globular
cluster (MKZR07). 
However, a high X-ray luminosity alone
does not guarantee a black hole source in extragalactic globular
clusters, as multiple neutron stars are a possibility in these
systems which are unresolved in X-ray data.
Variability has long been recognized as a key for distinguishing 
a black hole X-ray binary from multiple neutron stars in 
extragalactic globular clusters (e.g.\ Kalogera, King, \& Rasio 2004).
Therefore, the clear variability observed during the XMM
observation unambiguously indicates that the high X-ray luminosity 
in this source originates from a black hole (MKZR07).

	This object is the first and currently only confirmed 
black-hole X-ray binary in a globular cluster. As such it opens 
up new possibilities to address questions about the evolution of
black holes in dense stellar systems. It also provides a unique
opportunity to study in a nearby system how the energy from a 
highly luminous accreting black hole is fed back into its
surrounding medium in a dense stellar system. 
To begin considering these questions, this letter presents initial 
optical spectroscopy of this black-hole X-ray binary and the globular 
cluster in which it is found. We present the observations in
Section 2, describe the results of their analysis in section 3,
and discuss their implications in Section 4.

\section{Observations}

	As part of a study of the kinematics of the globular
cluster system of Virgo elliptical galaxy NGC~4472, we used the 
FLAMES multi-fiber spectrograph (Pasquini et al.\ 2004)
on the VLT/UT2 to obtain spectra 
of globular clusters over a wide area around NGC~4472 (Zepf et al.\ 2007).
The globular cluster targets were selected from the multicolor,
wide-field imaging study of Rhode \& Zepf (2001, hereafter RZ01). 
RZ01 used $BVR$ MOSIAC images from the Mayall 4-m to identify 
candidate globular clusters over an approximately $35' \times 35'$ 
field of view.
The spectroscopic data were obtained with FLAMES in its GIRAFFE/MEDUSA 
mode, which offers 130 fibers over a 25' diameter field of view.
By observing through four fiber setups, each slightly offset in
a different direction from the center of the galaxy, we were able 
to obtain spectra for nearly 400 globular clusters over a large 
range of distances from the center of the galaxy, with good 
azimuthal coverage, and many overlaps between pointings to provide 
checks on the accuracy of the radial velocities of the clusters.

The observations were carried out in service mode at the VLT/UT2
with two fiber setups observed in Jan 2004 and two observed
in Apr 2005. All of the data were obtained
with grating LR4, which covers the range 5015 \AA\ to 5831 \AA\
with a resolving power $\lambda/\Delta(\lambda)$ of 6000.
This range covers many stellar absorption features including strong
Mg and Fe lines and is therefore excellent for determining
the radial velocity of globular clusters. The fairly high
spectral resolution also results in small velocity 
uncertainties (see Bergond et al.\ 2006, Zepf et al.\ 2007). 
For studying possible emission lines from the black-hole X-ray binary,
the narrow wavelength range is imperfect. However,
it does provide a first look at the spectrum of the unique source,
and the spectral resolution corresponding to a velocity width
of about 50 \kms\ is useful for assessing whether any of the lines 
are broader than this value.

	The bright X-ray source XMMU 122939.7+075333 identified in the
XMM and Chandra images is located at the position of the globular
cluster candidate RZ 2109 in the Rhode \& Zepf 2001
cluster sample.  The astrometric matching of these
catalogs was described in MKZR07, with the agreement 
good to $< 0.4''$, a level similar to the positional
uncertainties in the catalogs themselves.
As discussed in MKZR07, this excellent agreement gives
a negligible probability ($ < 10^{-6}$) that the
X-ray source and globular cluster are not associated with
one another. This globular cluster was observed in two of 
the FLAMES fiber setups described above.
One of these was observed for a total of 4.3 hours in Jan 04,
and the other for 3.2 hours in Apr 05.
The reduction of the FLAMES multifiber spectra was carried
out in the same way as our previously published analysis
of similar data for globular clusters around NGC~3379 (Bergond et al.\ 2006).

Our standard analysis of the radial velocity of RZ 2109
from the cross-correlation of its absorption lines with stellar templates 
gives a radial velocity of $1477 \pm 7$ \kms\ for the Jan 04 
observations, and $1460 \pm 19$ \kms\ for the Apr 05 observations. 
We adopt a weighted average of $1475 \pm 7$ \kms as the best estimate 
of the globular
cluster radial velocity (in MKZR07 we used the $1477$ \kms\
of the observation with the smallest error). The agreement
between the two gives high confidence in this radial velocity,
and its value unambiguously identifies the object as member 
of the NGC~4472 globular cluster system, which has a mean 
radial velocity of 1018 \kms\ and a velocity dispersion 
of about 300 \kms\ (e.g.\ Zepf et al.\ 2000).

\section {Evidence for [OIII] 5007 emission}
To test for the presence of emission lines in the optical spectrum
of XMMU 122939.7+075333, we visually inspected the two independent
FLAMES spectra of the object described above. As shown in Figure 1,
both spectra show evidence for an emission line at a wavelength
corresponding to [OIII] 5007 at the radial velocity observed
for the host globular cluster. Although the line is only seen
at modest signal-to-noise in each spectrum, the fact that it 
is observed in the same place in the two independent spectra,
and that it is located precisely at the wavelength of [OIII] 5007
redshifted by the radial velocity of the cluster strongly
supports this identification of the line. More specifically,
we find the line center to be  5031.5 \AA\ in Jan 04 spectrum, 
while in Apr 05 spectrum, we find a line center of 5030.8 \AA\ 
with an uncertainty of several-tenths of an angstrom, based on 
both gaussian fits and simple mean wavelengths of the continuum 
subtracted emission line. 
We adopt $5031.2 \pm 0.3$ \AA\ as the best estimate
of the emission line center. Based on a zero redshift wavelength
of 5006.8 \AA\ for the [OIII] line, we then find a radial
velocity of the line of about 1462 \kms\ with an uncertainty 
of about 18 \kms, in excellent agreement
with 1475 \kms\ found for the stellar absorption lines of the
host globular cluster.

A second key feature of the line seen in each spectrum is
that it appears to be broader than the spectral resolution
of the data. Specifically, standard gaussian fits to the line
give a FWHM of $4 - 6$ \AA\  in both independent spectra.
This is much larger than the 0.9 \AA\ (4.5 pixels) FWHM of the 
effective instrumental resolution. 
The [OIII] 5007 line then is clearly intrinsically
broad, with a width of about $200 -350$ \kms. This measured
width does not rule out either a weaker, broader component,
or detailed structure within the line.
However, it does clearly indicate the overall [OIII] 5007 emission
is spread over at least a few hundred \kms.

This observed velocity width strongly implicates the black hole as 
the source of the energy driving the [OIII] 5007 emission. 
Without a constraint on the line width, the possibility would
remain that a planetary nebula in the globular cluster
is the source of the [OIII] emission. While these are
rare, there are two known cases among the hundreds of extragalactic
globular clusters with published spectroscopy (Bergond et al.\ 2006
and Pierce et al.\ 2006 for a NGC~3379 globular cluster, and 
Minniti \& Rejkuba 2002 for a NGC~5128 globular cluster). 
However, planetary nebulae have velocity widths 
of tens of \kms\ (Acker et al.\ 1992 and references therein),
and therefore the observed velocity width of several hundred \kms\
eliminates the possibility that the optical emission is
from a PNe. 

The luminosity of the emission line can also be estimated
from the spectra. Although flux standards were not
observed as part of this program, the target globular 
cluster has a known magnitude of V=21.0 (RZ01).
As a result, we can fix the flux in the continuum to be that
for a globular cluster with V=21.0, and use that
flux calibration to estimate the observed flux in the
[OIII] 5007 line. We can also check that the 
detailed flux distribution across the V band agrees
with that expected from models of stellar populations
(e.g.\ Bruzual \& Charlot 2003) like that of the globular
cluster and its known colors. 
It is worth noting that the uncertainty 
in the determination of the total counts in the emission
line is substantial, as described below, and thus the uncertainty 
in this flux calibration contributes negligibly
to the error budget of the line luminosity determination.

It is in this way that the flux calibration was estimated
for the two spectra shown in Figure 1.
An independent check on this calibration can be made by
comparing the observed signal to noise to that
expected given the cluster's known magnitude and
the observing conditions, based on the well characterized
throughput of the FLAMES instrument. We find good agreement
between these, providing further evidence that the flux
estimate is reasonable, and that the uncertainty in
the line luminosity is dominated by the measurement 
of the line itself. 

The [OIII] line luminosity can then be determined by measuring the 
total line flux in the calibrated spectrum, and adopting a distance 
to NGC~4472 of 16 Mpc (Macri et al.\ 1999). 
The most uncertain part of this procedure is determining the total
fux in the line, and in particular setting the level of the
continuum and the range of wavelengths over which to determine
the line flux. Our best estimate from both of our spectra in
Figure 1 is that L([OIII] 5007) is about $ 4 \times 10^{35}$ \ergss\ .
However, if the line is broader than we are able
to clearly discern in our data, we estimate this value
could be roughly twice as large and we also find it might
be possible to to fit the line with a flux that is smaller by $50\%$. 
Therefore, we attribute a factor of two uncertainty 
to our L([OIII] 5007) $\simeq 4 \times 10^{35}$ \ergss\ estimate.

\section {Discussion}

We have shown that optical spectroscopy reveals [OIII] 5007
emission associated with the black-hole X-ray binary system
in a globular cluster around the elliptical galaxy NGC~4472.
This [OIII] 5007 emission clearly establishes an interaction
between the X-ray binary and material around the black hole 
system. As such it also provides a potentially 
valuable testbed for understanding the relation between the 
binary evolution and accretion history of black holes and 
their interaction with the surrounding interstellar medium. 
Our spectra have also been able to establish the general 
velocity width of the [OIII] 5007 line. The several 
hundred \kms\ width of the line both demonstrates
the emission is associated with the black-hole X-ray binary
since no other cluster source could produce it, and also
shows the material may be driven out of the globular,
as the velocity width is much larger than the escape velocity 
of globular clusters.

[OIII] 5007 can be produced in principle by either collisional 
or photoionization. We consider each of these possibilities and 
their implications for the mass of the black hole. In the collisional
case, a strong wind is generated by the black-hole X-ray binary, 
which then drives a shock front into the interstellar medium 
of the globular cluster. We can write standard solutions 
for the radius $R$ of this shock front at time $t$ as
$R(t) = 28 [(\dot{M}/10^{-6}$ 
M$_{\odot}$/yr)$(v_0/20000$ \kms)$^2 (0.1 {\rm cm}^{-3}/n_0)]^{1/5} 
(t/10^{5} \rm{yr})^{3/5}$ pc 
(e.g.\ Castor, McCray, \& Weaver 1975), 
and the current velocity as $V(t) = 3/5 R(t)/t $,
where $\dot{M}$ is the mass loss rate in the wind, $v_0$ the
initial velocity, and $n_0$ the density of the interstellar medium.
We now consider what likely values for $n_0$, $\dot{M}$, $v_0$, and $t$ 
are and how these might be constrained by current and future
observations.

The density of the interstellar medium of globular clusters, $n_0$,
is only well-known in a few cases. The best constraint is
probably for 47 Tuc, for which the gradient in pulsar dispersion 
measurements with radial position in the cluster gives a 
free-electron density of $0.067 \pm 0.015$ cm$^{-3}$ (Freire et al.\ 2003).
Similar studies of pulsar dispersion measurements give a
similar or slightly higher density in M15 (Freire et al.\ 2001),
and a recent study may have detected somewhat
more neutral hydrogen in M15 (van Loon et al.\ 2006).
Most other studies of globular clusters have only been able to
place upper limits which are typically well above the adopted value. 
That there would be some interstellar medium in a globular cluster 
seems natural, as mass loss from
the cluster stars provides a continual source of interstellar
material. Even if many processes cause clusters
to also lose ISM, the velocities with which this mass is
driven out would have to be extraordinarily high or
else the steady state density of the ISM will be generally
similar to the fiducial value adopted here
(Pfahl \& Rappaport 2001). 

The radius and current velocity of the shocked region also depend
on the kinetic luminosity in the wind, $\dot{M} v_0^2$.
In the context of super-Eddington accretion onto a stellar mass 
black hole, this should be close to the Eddington luminosity 
of the source, and should be independent of the actual accretion 
rate, providing it is well above 
the accretion rate require to yield the Eddington luminosity for 
an accretion efficiency of 0.1$c^2$ (Begelman, King \& Pringle 2006).
The observed X-ray luminosity of $L_X=4\times10^{39}$ ergs/sec and 
and inner disk radius of $\sim 4000$ km inferred from the X-ray data 
(MKZR07) can be well fit by this model with a 10 $M_\odot$ black hole 
accreting at about 40 times the rate required to give the Eddington 
luminosity.

In this case, the predicted initial wind velocity $v_0$ is 20000 km/sec 
and $\dot{M}$ is $8 \times 10^{-6}$ M$_{\odot}$. These then give 
a timescale $t$ of about $2 \times 10^5$ years for the bubble to 
expand to the point where the one-dimensional velocity of the outflow 
is about 150 km/sec. The corresponding size of the bubble is about
40 pc, which is within the tidal radius expected for a fairly
massive outer globular cluster like RZ 2109. 
Taken at face value, a model in which $\dot{M}$ is constant
over the above timescale loses somewhat more mass than
the 1 M$_{\odot}$ upper limit for a donor star in the globular
cluster. However, the detailed application of this model 
involves many uncertainties including the level of isotropy 
of the accretion flow, uncertainties in the radii inferred 
from X-ray spectroscopy, and inhomogeneity in the ISM along 
with the effect of input of additional material from ongoing 
stellar mass loss in the globular cluster. 
It would then seem a wind-driven system 
with a constant $\dot{M}$ remains feasible. 

Alternatively, if the system experiences periods of quiescence, 
the total mass lost and kinetic power are less than
calculated assuming these are constant at the above values.
This is because the timescale is the time since the
object first became an energetic binary, while $\dot{M}$ and
thus the kinetic luminosity are the time averaged values since 
this time, and are thus reduced directly by the duty cycle. 
In either case, the calculation
suggests that a scenario in which a super-Eddington stellar 
mass black hole driving a strong wind through the interstellar 
medium of the globular cluster may be able to produce the observed 
[OIII] 5007 emission line. The X-ray data also allow the possibility
that the accretor is an intermediate-mass black hole (IMBH)
emitting at sub-Eddington luminosities (MKZR07). However,
a strong wind is not expected in the IMBH sub-Eddington case 
(Proga 2007 references therein), so if the [OIII] 5007
emission is from shocks, then it would seem the black hole
has a stellar mass.

We now consider the possibility that the [OIII] 5007 emission line
is photoionized. Because the velocity width of the line is significantly 
greater than the virial velocity of the globular cluster, if the
line is photoionized, it must be produced near the black hole. 
For a $10^3$ M$_{\odot}$ black hole, the velocity width indicates 
the line must come from about $\sim 7 \times 10^{14}$ cm away 
from the black hole. This may be possible, although it may prove
difficult to have enough mass at the right distance from 
the black hole to produce the [OIII] 5007 line luminosity at the 
observed velocities. For completeness
we note that for a stellar mass black hole the observed
[OIII] 5007 seems unlikely to be due to photoionization.
The broad lines observed in Galactic black-hole
X-ray binaries are typically permitted 
lines such as the Balmer lines and He II 4686, as might be expected 
due to the high densities at the small distances from the central 
source required to match the velocities. 
In contrast, as described above 
a shock is expected if the accretor is a stellar mass black hole 
and can produce roughly the observed [OIII] 5007 emission. 
A final alternative is it may be possible for a stellar
mass black hole to drive a wind as described earlier, but
for photoionization of the wind material by the central source
to play a significant role in producing [OIII] 5007.

This analysis shows that undertanding the physical process
which produces the [OIII] 5007 emission can place constraints 
on the mass of the black hole.
Determining the mass of the black hole has important implications 
beyond understanding this individual system. 
Specifically, it would seem very unlikely to find a stellar mass
black-hole X-ray binary in a globular cluster with an
intermediate mass black hole. This is because most black holes
are believed to be either dynamically ejected (Kulkarni, Hut, \& 
McMillan 1993, Sigurdsson \& Hernquist 1993) or 
merge to form a IMBH (Miller \& Hamilton 2002). Any surviving stellar 
mass black hole in a cluster
with an IMBH would have to reside in the outer regions 
of the cluster to avoid either of these fates. However,
a location in the outer region of a globular cluster
with its low stellar density is not conducive to the
stellar dynamical interactions that make close, accreting binaries
with black hole primaries in globular clusters.
Therefore, it appears that if it can be shown the black hole
must have a stellar mass, then this globular cluster would 
be very unlikely to host an IMBH. This would then 
demonstrate that the $M_{BH} - \sigma$ relation for larger mass 
stellar systems does not hold for all stellar systems at globular 
cluster masses.

There are several future observations that will help constrain
the physical parameters of this interesting system.
One of these is expanding the wavelength coverage beyond the narrow region
currently available. This will provide line ratio diagnostics
such as [SII]/H$\alpha$ and [OIII]/H$\beta$  
to determine the relative importance of collisional and photoionization
processes. As described above, this will have immediate implications for
the mass of the accreting black hole, as the currently observed
[OIII] 5007 would seem to require photoionization for an IMBH,
while shock ionization works well in the case of a stellar mass
black hole. Wider wavelength coverage would also lead to
constraints on the density structure in the cluster 
interstellar medium, and possibly assessing the total power and 
lifetime of the system.

A key parameter that may be constrained by future
observations is the radius at which the [OIII] 5007 emission
originates. We note that at the 16 Mpc distance of NGC~4472, 
$1''$ corresponds to 77 pc, so images with much higher
spatial resolution than this, such as from HST, may be
able to resolve the nebulae and determine its size
if it originates in shocks. Correspondingly, a tight
upper limit would suggest photoionization, as the
spatial scale in this case is orders of magnitude smaller
and unresolvable. Moreover, if the line emission originates in shocks,
resolving it would allow the size and velocity width
to be combined to directly give the timescale, and the total kinetic 
luminosity would then follow. This would also
provide a constraint on whether the accretion has
been constant or intermittent, since it is the time averaged 
kinetic luminosity that matters for the shock radius while
the timescale that matters is the total lifetime of the
system. 

\acknowledgments

SEZ and AK acknowledge support for this work from
XMM grant NNG04GF54G, NSF award AST-0406891 (SEZ), and
NASA-LTSA grant NAG5-12975 (AK),
GB is supported at the IAA/CSIC by an I3P contract (I3P-PC2005-F) 
funded by the European Social Fund, with additional support by 
DGI grant AYA 2005-07516-C02-01 and the Junta de Andaluc\'\i a,
and KLR acknowledges support from a NSF Astronomy and Astrophysics 
Postdoctoral Fellowship under award AST-0302095. 
The authors acknowledge useful discussions with 
Phil Charles, Mike Eracleous, Andrew King, Dave Russell, and Mark Voit.
We also thank the referee for a prompt and effective report.

\clearpage

\begin{figure}
\plotone{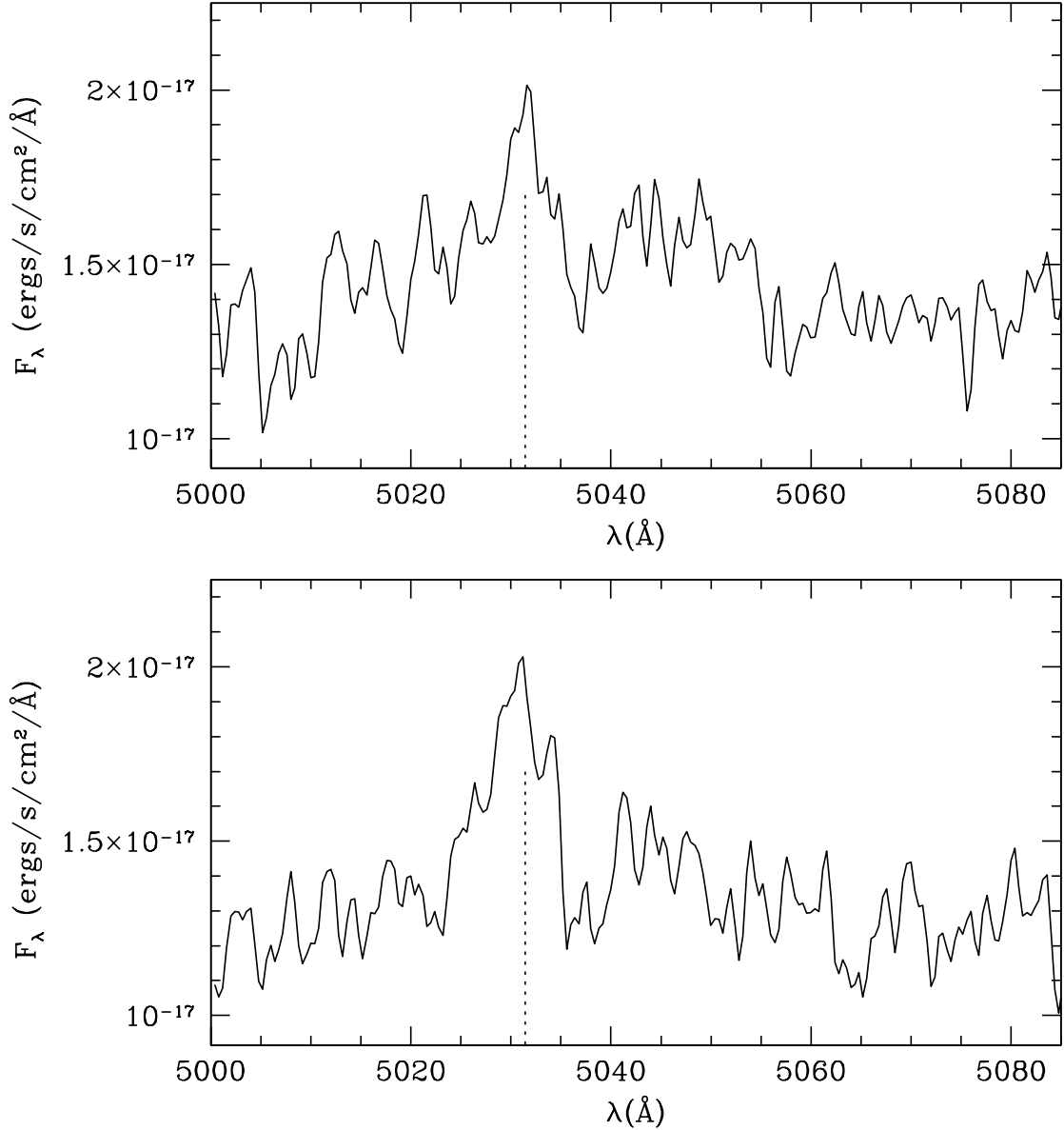}
\caption{Plots of the optical spectra of the globular cluster 
shown to host a black hole X-ray binary by MKZR07. Both of the
two independent fiber spectra from FLAMES on the VLT show 
emission with a central wavelength of 5031.2 \AA\ 
with a uncertainty of several tenths of an angstrom, 
and a line width of about $4 - 6$ \AA\ . The observed wavelength
strongly supports the identification of this line as 
[OIII] 5007 from the globular cluster, which has a 
radial velocity of the 1475 \kms\ determined from 
its stellar absorption lines. The dotted line denotes the
wavelength expected at [OIII] 5007 at this redshift.
The $4 - 6$ \AA\ width of the [OIII] 5007 line is clearly
broader than the $0.9$ \AA\ spectral resolution of the data. 
This width corresponds to several hundred \kms, much larger than
any planetary nebulae, and implicates the black-hole
X-ray binary as the source of the energy driving this emission.
The plotted data have been smoothed by the instrumental
resolution, and the calibration of the flux given on the y-axis 
is described in the text. 
\label{fig1}
}

\end{figure}

\end{document}